
\magnification=\magstep1
\voffset=-0.8 true in


\newcount\equationno      \equationno=0
\newtoks\chapterno \xdef\chapterno{}
\def\eqn{\eqno\eqname}
\def\eqname#1{\global \advance \equationno by 1 \relax
\xdef#1{{\noexpand{\rm}(\chapterno\number\equationno)}}#1}



\def\msun{{\rm M}_\odot}
\def\kms{{\rm km\;s}^{-1}}

\centerline{\bf BARYONIC CONTENT OF GALACTIC HALOS AND CONSTRAINTS}
\centerline{\bf ON MODELS FOR STRUCTURE FORMATION}
\vskip 2 true cm
\centerline{\bf T.Padmanabhan$^1$ and K.Subramanian$^{2*}$ }\footnote{}{*The
ordering of the names of the authors was determined by tossing a
coin, which is a somewhat simpler procedure than reading T.S.Eliot's
poems}
\vskip 1 true cm
\centerline{$^1$Inter University Center for Astronomy and Astrophysics,}
\centerline{Post Bag 4,Ganeshkhind,Pune 411 007,India}
\smallskip
\centerline{$^2$ National Center for Radio Astrophysics, TIFR}
\centerline{Post Bag 3,Ganeshkhind,Pune 411 007,India}
\vskip 1 true cm
\centerline{IUCAA-23/93; sept,93; submitted for publication}
\vskip 2 true cm
\noindent
{\bf
The recent detection of microlensing of stars of LMC by compact objects
in the halo of our galaxy$^{1-2}$ suggests that our galaxy is surrounded by a
non-luminous halo made of compact objects with mass of about $(0.03-0.5)
\msun$. The rate of detection could be consistent$^{3-4}$
with the assumption that these
halo objects are distributed with a softened isothermal profile with a core
radius of $(2-8)$Kpc and asymptotic circular velocity of $220\kms$. Taken in
isolation, this observation is consistent with a universe having only baryonic
dark matter (BDM, hereafter) contributing $\Omega_b=\Omega_{total}\simeq0.06$.
Such a model, however, will violently contradict several other large scale
observations, notably the COBE-DMR results. The simplest way to reconcile the
microlensing observations with such constraints is to assume that galaxies like
ours are surrounded by both BDM and non-baryonic dark matter (NBDM, hereafter).
A model with a single component for NBDM with, say,
$\Omega_b\simeq0.06,\Omega_{cdm}\simeq0.94$, is also ruled out
if we demand that: (i) at least thirty percent of the dark matter density
within $100$ kpc is baryonic and (ii) galactic structures should have collapsed
by redshift of $z=1$. If further microlensing observations suggest that
half  or more of the dark matter within $100$ kpc is baryonic, then we are led
to the powerful
constraint that the maximum value of $\Omega_{dm}$, contributed by NBDM
clustered at galactic scales, is about $\Omega_{max}\simeq0.29$. Hence if we
demand that $\Omega_{tot}=1$, then about 70 percent of dark matter must be
distributed smoothly over galactic scales. Models with C+HDM cannot satisfy
this
constraint but $\Lambda$+CDM models are still viable.
 Even in such a hy-hybrid model it is not
clear whether one can consistently explain the abundance of quasars and
absorption systems.
}

Several experiments are underway at present$^5$ to detect microlensing
of stars in the Large Magellanic Cloud by compact objects (MACHOs)
which could exist in the dark halo of our galaxy. If this dark halo
is modelled as softened isothermal sphere made of MACHOs,
with a core radius of
$(2-8)$ kpc and a rotation velocity of $220 \kms$, then the
expected optical depth to microlensing is$^{3-4}$ about $5 \times 10^{-7}$.
Recently 3 such possible  microlensing events have been reported$^{1-2}$
which could be consistent with this optical depth. The time scale of the
flux variation of the background stars
indicates a MACHO mass in the range $(0.03-0.5)\msun$.
This observation, taken in isolation, is consistent with a universe
having only baryonic dark matter with $\Omega_b \simeq 0.06$ - a value which
will also satisfy the bounds from primordial nucleosynthesis if $h=0.5$.
However,
a model with $\Omega_{total}=\Omega_b=0.06$ will produce too large$^{6}$ a
microwave anisotropy to be consistent with COBE-DMR observations. (Adiabatic
fluctuations will be worse than isocurvature perturbations in this regard
but both will be ruled out). Such a model will also have difficulty in
explaining several other large scale observations and the virial mass estimates
of
clusters.

One is, therefore, led to study models in which galctic halos contain both
baryonic and non-baryonic dark matter. Let us consider the simplest of such
models in which BDM and cold dark matter (CDM) halos around a galaxy are
described by the density
profiles
$$\rho_{bdm}={\rho_b\over1+(r/r_b)^2};\quad
\rho_{cdm}={\rho_c\over1+(r/r_c)^2}\eqn\denpro$$
We shall take a baryonic core radius of $r_b=2$ kpc, but keep the other
parameters unspecified at this stage. The square of the rotational velocity
$u^2(r)=u_c^2(r)+
u_b^2(r)$ at any radius is the sum of the contributions from CDM and BDM with,
for example,
$$u_b^2(r)=v_b^2\left[1-{tan^{-1}(r/r_b)\over (r/r_b)}\right]\eqn\qq$$
where $v_b^2=4\pi G\rho_b r_b^2$ and with a similar expression for $u_c^2(r)$.
We shall define a parameter $\lambda$ by  $v_b=220\lambda
\kms$ so that $\lambda^2$ represents the baryonic contribution to density
asymptotically. The mass of the BDM within $100$ kpc can be easily found by
integrating \denpro\ and we get $M_{bdm}(100 kpc)\simeq \lambda^2\times10^{12}
\msun$. If we take the CDM halo attched to the galaxy to be $\mu$ times more
massive, we find the CDM contribution to halo mass to be $M_{cdm}=\mu\lambda^2
\times 10^{12}\msun$. If such a halo has collapsed and virialised by
a redshift $z$, then we can relate the $z,M_{cdm}$ and $v_c$ using the
spherical top hat model. This relation$^7$ gives
$$v_c=81.65\kms (1+z)^{1/2}\lambda^{2/3}\mu^{1/3} \eqn\vcdm$$
Asymtotically, we need to obtain a flat rotation velocity of $220\kms$ for our
galaxy$^8$. Using
\vcdm\ and setting $v_b=220\lambda\kms$ in the relation $v^2=v_c^2+
v_b^2=(220\kms)^2$ at large $r$, we get $\lambda^2+0.138\lambda^{4/3}
\mu^{2/3}(1+z)=1$ or, equivalently,
$$\mu=19.5{(1-\lambda^2)^{3/2}\over\lambda^2(1+z)^{3/2}}\eqn\const$$
This condition relates $\mu=(M_{cdm}/M_{bdm})$ to the redshift of formation
($z$)
and the contribution of BDM to rotation velocity ($\lambda$). The limit
$\lambda=1$
corresponds to a purely BDM model with $\mu=0$. We are interested in a model
with $\Omega_{tot}=1,\Omega_B\simeq0.06,\Omega_{cdm}\simeq0.94$ for which
$\mu=\Omega_{cdm}/\Omega_{bdm}=0.94/0.06=15.67.$ Even if we assume that
as much as half
the mass within $100$ kpc is nonbaryonic (so that $\lambda^2=0.5$), we only get
$\mu=13.8$ for $z=0$ and $\mu=4.88$ for $z=1$. One has to lower the value of
$\lambda^2$ to about $0.273$ to get $\mu=15.67$ for $z=1$. In other words,
this model can be ruled out if more than about 27 percent of the dark matter
within $100$ kpc is in baryonic MACHOs.

The precise bound on the amount of NBDM that can exist within
$100$ kpc depends on the detailed statitics of the microlensing events and as
the statistics improves, we will have tighter bounds on nonbaryonic
contribution.
(that is, the minimum allowed value for $\lambda$ will increase). We have also
tried to see whether it is possible to accommodate CDM and BDM by increasing
$r_c$ and decreasing $\rho_c$. We find that it is not possible to have flat
rotation curve with the correct asymptotic value (since for large $r_c$, CDM
contributes a $u_c(r)\propto r$)
 for any combination of
parameters which satisfy the other constraints. It seems
that the simplest model with both CDM and BDM coexisting in the halo
is ruled out if: (a) at least half the dark matter within $100$kpc is
contributed
by BDM and (b) galactic halos have collapsed at least by the redshift of $z=1$.

If the second condition is relaxed, one can marginally satisfy the first
condition
with $\lambda^2=0.48$. Such a model still faces other difficulties. In such
a scenario each galaxy is associated with $M_{bdm}\simeq 4.8\times10^{11}\msun$
and $M_{cdm}\simeq 7.2\times10^{12}\msun$ which is about an order of magnitude
higher than the conventional masses associated with galxies. Correspondingly,
the masses associated with clusters will be one order of magnitude higher.
Clusters
will therefore contribute an amount $\Omega_{clus}\simeq0.09(M_{clus}/5\times
10^{15}h^{-1}\msun)$ to the closure density. From the Press-Schecter
analysis$^{9-10}$,
one can relate  the linear density contrast $\sigma(M)$ at $M$ to the
fractional contribution to the density, $\Omega(M)$, from masses higher than
$M$ at a redshift $z$. The relation is
$$\Omega(M)=erfc\left[\delta_c(1+z)\over\sqrt{2}\Delta(M)\right]\eqn\omab$$
where $\delta_c=1.68$ and $erfc(x)$ is the complementary error function. To get
$\Omega=0.09$ at $z=0$ at cluster scales we need $\Delta(M_{clus})\simeq 1$.
This condition is difficult to achieve in standard models without making the
slope of galaxy-galaxy correlation function too high.

These difficulties are independent of the processes$^{11-12}$ which may have
led to the
formation of baryonic compact objects in our halo. Of course, astrophysical
considerations regarding the formation of brown dwarfs etc. will impose other
constraints
on these models. However, the arguments given above appear to be more robust
and free from astrophysical uncertainties.

If one insists that $\Omega_{tot}=1$ and $\Omega_b=0.06$, then it seems
necessary
to distribute the NBDM in two components, one clustered at galactic scales
and another which has a smoother distribution. Assuming that at least half
the matter within $100$kpc is baryonic(i.e, $\lambda^2=0.5$) and that galactic
structures have collapsed by $z=1$,
we find from \const\ that $\mu=4.88$. Since the maximum value for $\Omega_b$
permitted by nucleosynthesis is about 0.06, we conclude that the clustered
non-baryonic
mass in galactic scales can only be $\Omega_{max}=\mu\Omega_{bmax}\simeq
0.29$. Thus if we want a flat universe, nearly 70 percent of the dark matter
must be unclustered. Two component dark matter models with C+HDM or
$\Lambda$+CDM were extensively investigated recently$^{13-17}$ since they were
in better agreement
with large scale observations and COBE. Of these, H+CDM models have$^{16}$
$\Omega_{cdm}\simeq 0.7$ and are not compatible with our constraints. The
$\Lambda$+CDM models can be made marginally consistent$^{17}$ with
$\Omega_{cdm}
\simeq (0.2-0.3)$ and could be viable. This model needs to be studied more
carefully
especialy as regards abundance of quasars and absorption systems.$^{18}$
Even this model will be ruled out if the improved analysis
of lensing events shows that significant fraction of halo dark matter is
baryonic.

We have assumed throughout the discussion that machos are made of baryons.
If they are condensates of more  exotic elementary particles or primordial
blackholes, then it is possible that MACHOs can mimic some properties of
CDM and the conclusions could be different. However, the mass scale of
$0.1\msun$ enters the hubble radius when $T\simeq 1.1 MeV$. Since no
unusal physics seem to occur at this energy, it seems somewhat unlikely that
exotic
objects with this mass scale can be formed with sufficient abundance to provide
closure density.

\centerline{\bf Aknowledgements}
\noindent
K.S thanks V.Sahni,B.S.Satyaprakash,T.P.Singh and R.P. Sinha for communicating
information on the discovery of MACHOs.
\beginsection\centerline{\bf References}

\item{1.} C.Alcock et al., (1993) CfPA preprint 93-30; to appear in Nature
\item{2.} J.Rich (1993) as cited by C.Alcock et al., ref.1
\item{3.} K.Griest (1991), Ap.J., {\bf 366}, 412.
\item{4.} B.Paczynski (1986), Ap.J, {\bf 304},1.
\item{5.} For a review, see e.g, D.P.Bennett in {\it After the first three
minutes}, eds. S.S.Holt et al., (AIP, New York, 1991), p.446.
\item{6.} J.R.Bond (1988) in {\it The early universe}, eds. W.G.Unruh and
G.W.Semenoff, (Reidel, Dordrecht)p.283
\item{7.} T.Padmanabhan (1993) {\it Structure formation in the universe},
(Cambridge University Press); eq.(8.45).
\item{8.} J.Binney and S.Tremaine (1987),{\it Galactic Dynamics}, (Princeton
University Press)
\item{9.} W.H.Press and P.L.Schecter, (1974), Ap.J., {\bf 187},425.
\item{10.} S.D.M.White et al.,(1993), MNRAS, {\bf 262},1023.
\item{11.} K.M.Ashman., B. J. Carr (1989), MNRAS, {\bf 234}, 219.
\item{12.} A.C.Fabian (1990), in {\it Baryonic dark matter}, eds. D.Lynden-Bell
and G.Gilmore (Kluwer Academic Publishers), p.195.
\item{13.}A.N.Taylor and M.Rowan-Robinson (1992), Nature, {\bf 359},396
\item{14.}M.Davis et al.,(1992), Nature, {\bf 359}, 393
\item{15.}A.Klypin et al.,(1993) preprint SCIPP 92/52 (submitted to Ap.J)
\item{16.}D.Y.Pogosyan, A.A.Starobinsky (1993) IOA preprint
\item{17.}L.Kofman et al.,(1993), CITA preprint.
\item{18.} K. Subramanian and T. Padmanabhan (1993) in preparation.

\end